\documentclass[final,3p,times,twocolumn]{elsarticle}
\usepackage{graphics}
\usepackage{graphicx}
\usepackage[section] {placeins}
\usepackage{epsfig}
\usepackage{pbox}
\usepackage{amssymb}
\usepackage{amsthm}
\usepackage{amsmath}
\usepackage{lineno}
\usepackage{breqn}
\usepackage{amssymb,amsmath}
\usepackage{setspace}
\usepackage{graphicx}
\usepackage{epsfig}
\usepackage{makecell}
\usepackage{pbox}
\usepackage{rotating}
\usepackage{multicol}
\usepackage{multirow}
\usepackage{lscape} 
\usepackage{booktabs,caption,fixltx2e}
\usepackage[flushleft]{threeparttable}
\usepackage{tablefootnote}
\usepackage{longtable}
\usepackage{amsfonts}
\usepackage{float}
\usepackage{booktabs}
\usepackage{graphicx}
\usepackage{array}
\usepackage{blindtext}
\usepackage{lipsum}
\usepackage{enumitem}

\begin{document}

\begin{frontmatter}

\title{Multi Anode Photomultiplier Tube Reliability Assessment for the JEM-EUSO Space Mission}

\author[Affil1]{\corref{cor1}H. Prieto-Alfonso}
\author[Affil1]{L. Del Peral}
\author[Affil2,Affil3]{M. Casolino}
\author[Affil2]{K. Tsuno}
\author[Affil4]{T. Ebisuzaki}
\author[Affil1]{M.D. Rodr\'iguez Fr\'ias} 
 \author{for the JEM-EUSO collaboration} 
\address[Affil1]{SPace and AStroparticle (SPAS) Group, UAH, Madrid, Spain}
\address[Affil2]{EUSO Team. RIKEN, Japan}
\address[Affil3]{INFN Roma Tor Vergata, Roma, Italy}
\address[Affil4]{Computational Astrophysics Laboratory. RIKEN, Wako, Japan}

\cortext[cor1]{Corresponding Author: H. Prieto-Alfonso, University of Alcal\'a, Alcal\'a de Henares, 28871, Madrid, Spain; Email:hector.prietoa@uah.es.}

\begin{abstract}
Reliability assessment is concerned with the analysis of devices and systems whose individual components are prone to fail. This reliability analysis documents the process and results of reliability determination of the JEM-EUSO Photomultiplier tube component using the methods 217 Plus, Quantum efficiency degradation and radiation hardness assurance. In conclusion, the levels of damage suffered by the PMTs which comprise the focal surface of JEM-EUSO Space Telescope, are acceptable. The results show as well that the greatest contribution to the failure is due to radiation SET. The guaranteed performance of this equipment is a 99.45\%, an accepted value of reliability thus fulfilling the objectives and technological challenges of JEM-EUSO. It should be noted that the reliability values ​​of the Standard 217Plus, despite being a standard improved, and an updated version of MIL-HDBK-217 Plus does not have sections that include the analysis of radiation of space electronic equipment. The recommendation from this study is the inclusion of all real failure effects may face an electronic, mechanical, optical space equipment. 
\end{abstract}


\end{frontmatter}
%\linenumbers

\section{Introduction}

\begin{figure}
\begin{center}
\includegraphics[scale=0.75]{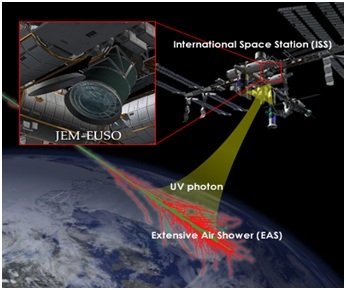} 
\end{center}
\caption{JEM-EUSO telescope onboard the International Space Station \cite{jemeusoweb}}
\label{jemeusoiss}
\end{figure}

JEM-EUSO \cite{jemeuso,jemeusoweb} is a large imaging telescope designed to study the Ultra-High Energy Cosmic Rays (UHECR) at energies above \textbf{$10^{20}$} eV, as represented by Figure \ref{jemeusoiss}. Looking downward the Earth from the International Space Station (ISS) it will detect such cosmic ray particles observing the UV light generated by Extensive Air Showers (EAS) while the UHECRs develop in the atmosphere. The scientific objectives of the mission include charged particle Astronomy and a promising branch of Astrophysics, called Astroparticle Physics, with the aim of extending the measurement of the energy spectra of the cosmic radiation beyond the Greisen-Zatsepin-Kuzmin (GZK) effect \cite{greisen}, together with the detection of Extremely High Energy Gamma Rays (EHEGR) and Extremely High Energy Neutrinos (EHEN). 

JEM-EUSO is being designed to operate for 5 years onboard the ISS orbiting in a Low Earth Orbit (LEO) around the Earth at an altitude of about 400 km. As for any mission to be operated in space, JEM-EUSO must comply with specific requirements, i.e. high radiation doses, unaccessibility and remote controlled operations. That is why the reliability analysis and radiation hardness assurance is extremely important in order to determine the tolerance and redundacy requirements within the system as previous studies show in case of Field Programmable Gate Array (FPGA) which also applies to PMTs \cite{Prieto, Prieto2, Prieto3}.

The design and the construction of the JEM-EUSO telescope is a real technical challenge, as it involves the use of new technologies from the laboratories of both industrial and research centers in areas as diverse as large optical and accurate Fresnel lenses, a novel technique of photo detection highly sensitive with very accurate resolution, and very innovative analog and digital electronics as well.

% Figure I ------> Ok!

The paper is organized as follows. First of all, we have explained  the considerations we have taken into account to make the reliability analysis, then a brief explanation of possible failures of PMT are shown in section 3, along with the elements that compose the PMT in section 4. In section 5 the reliability analysis of the PMT has been performed and finally in section 6 the conclusions and results of the analysis are presented.

\section{Technical Specifications of JEM-EUSO PMTs}

The focal surface is a spherical curved surface, its area amounts to about 4.5 m$^2$ and it is covered with about 5,000 Multi-Anode PhotoMultiplier Tubes (MAPMT) Hamamatsu R11265-03-M64 MOD: MAPMT designed for this space mission. The focal surface detector consists of Photo-Detector Modules (PDMs), each of which consists of 9 Elementary Cells (ECs). The EC implements 4 units of MAPMTs. Therefore, about 1,233 ECs or about 137 PDMs are arranged on the whole focal surface with 384,000 pixels \cite{Hamamatsu}.

Standards that describe PMTs reliability have not been developed so far, however we have considered the case of Vacuum Tubes (VT) and Integrated Circuits (IC) because of the similarity among them and PMTs. This has been done in order to develop a model for the PMT electronics behaviour taking as reference the most similar element in terms of functionality.

This study of the reliability of the PMT is a statistical study, not a performance study, referred to the failure rates of the PMT under the differents mechanism of failure, such a cracks in the envelope, electromigration, and multiple electronic behaviour such as short-circuits. In any case, this paper never tries to simulate the reliability of the performance of the PMT, which mean the reliability of the transmission of the signal, quality of the signal, neither the degradation of the performance during operation of the PMT (Background events), these events can be studied in other papers. In the another hand, this study will consider the PMT as a semiconductor, since most of the element are composed by alkali materials, considerations previously made by different studies including theory of the principal manufacturer of PMT as Hamamatsu \cite{HamaSemi}. in such a way we will proceed with this study taking into account the considerations also made by previous studies and theories \cite{Semiconductor1, Semiconductor2, Semiconductor3}.

If we try to compare PMT with ICs, the only reason we consider their relationship is because the IC and PMT in a reliability model are considered as system themself, however, according to this assumption, the IC is the closest electronic component to the PMT in the reliability standards we have so far \cite{PMTa,IC}. Table \ref{tab:pmtvsic} shows main differences among PMTs, VTs and ICs .

% Table I --------> Ok!
\begin{scriptsize}
\begin{table}[htb!]
\captionsetup{font=normalsize, labelfont=bf, labelsep=period}
    \caption{Differences among PMTs, VTs and ICs devices.}
    \centering
    \resizebox{8cm}{!} {
\begin{tabular}{c c c c r}
\toprule
\textbf{Element}&\textbf{Material} &\textbf{Feature Size}&\textbf{Physical Conditions}&\textbf{Operating Voltage (V)}\\
\midrule
\multirow{2}{*}{}&Anode&&2"$<$Size$\leq$30"&\\PMT&Cathode:(AgOCs,(SbKcS)&mm&Affected by Vibration& 500$<$ V $\leq$3000\\&Dynode:(AgMg),(CuBe)&&Affected by Shock&\\&&&Affected by Light&\\
\cmidrule{2-5}
\multirow{2}{*}{}Vacuum&Anode:Ni(Nickel)&&1.5"$<$Size$\leq$25"&\\Tube&Cathode:W (Tungsten)&mm&Affected By Vibration& 10 $<$ V $\leq$15000\\& Grid: Ni(Nickel)&&Affected By Shock&\\
\cmidrule{2-5}
\multirow{2}{*}{}&&&1.5"$<$Size$\leq$25"&\\Integrated&SiO$_2$(Silicon Dioxide)&nm&Vibration Resistance& 2.3 $<$ V $\leq$5\\Circuit&&&Shocking Resistance&\\&&&Affected by temperature&\\
\bottomrule
\end{tabular}
    }
    \label{tab:pmtvsic}
\end{table}
\end{scriptsize}

The PMTs used in JEM-EUSO have been specially designed and manufactured by Hamamatsu and are intended for this space mission. With the aim to improve the technical specification of the PMT for the JEM-EUSO Space telescope, Hamamatsu, the Japanese company, has developed dedicated PMTs according to the JEM-EUSO Space requirements. Tables \ref{PMT Characteristics} \& \ref{PMT Characteristics3} show technical characteristics of the PMT, with a spectral window in the UV enough broad to observe both the fluorescence  as Cherenkov light generated in the EAS. A gain ratio of  $10^6$ added to a cathode sensitivity of 90 $\mu $A/lm allows the detection of low light levels with a very low dark current of 0.4 nA. The PMT is operative in a wide temperature range. Table \ref{PMT Characteristics4} lists nominal voltage ratios for the dynode of the PMTs.

%Table II --------> Ok!
\begin{center}
\begin{table}[htb!]
\captionsetup{font=normalsize, labelfont=bf, labelsep=period}
    \caption{Technical specifications of the JEM-EUSO PMT \cite{PMThamamatsu}}
    \centering
    \resizebox{8cm}{!} {
\begin{tabular}{l|l}
\hline
Parameter & Description \\
\hline
Spectral Response Range&185 to 650 nm\\
Window material/Thickness&Ultra Violet glass/0.8 mm\\
Photocathode Material  & Bialkali  \\
Photocathode minimun effective area& 23$\times$23  mm$^2$ \\
Dynode structure & Metal channel Dynode \\
Number of stages & 12  \\
Weight & 27 g  \\
Operating ambient temperature&-30 to $+50^\circ$C\\
Storage temperature&-30 to $+50^\circ$C\\
Voltage Supply (anode-cathode) & 1100 VDC \\
Average anode current & 0.1 mA \\
\hline
\end{tabular}
    }
    \label{PMT Characteristics}
\end{table}
\end{center}

%Table III --------> Ok!
\begin{table}[htb!]
\captionsetup{font=normalsize, labelfont=bf, labelsep=period}
    \caption{Electronic characteristics of the JEM-EUSO PMT  at $25^{\circ}$C \cite{PMThamamatsu}}
    \centering
    \resizebox{8cm}{!} {
\begin{tabular}{l|l|r|r|r|r}
	\hline
	\multicolumn{2}{c|}{Parameter}&Min&Typ.&Max.&Unit\\
	\hline
Cathode Sensitivity&Luminous (2856K)&70&90&-&$\mu$A/lm\\
	\hline
Anode Sensitivity&Luminous (2856K)&-&90&-&A/lm\\
	\hline
\multicolumn{2}{l|}{Gain (Current Amplification)}&-& 1$\times10^6$&-&-\\
	\hline
\multicolumn{2}{l|}{\pbox{20cm}{Anode Dark Current (Each Anode)\\ (After 30 min. storage in darkness)}}&-& 0.4&4&nA\\	
\hline
Cathode Sensitivity&\pbox{20cm}{Anode Pulse Rise Time\\ Anode Transit Time} &\pbox{20cm}{-\\-}&\pbox{20cm}{0.6\\5.1}&\pbox{20cm}{-\\-}&\pbox{20cm}{ns\\ns}\\
\hline
\multicolumn{2}{l|}{Uniformity Ratio Between Anodes}&N/A&1:3&1:5&N/A\\
\hline
\pbox{20cm}{Pulse Linearity\\(Each Anode)}&\pbox{20cm}{$\pm$2\% Deviation\\$\pm$5\% Deviation} &\pbox{20cm}{-\\-}&\pbox{20cm}{0.2\\0.4}&\pbox{20cm}{-\\-}&\pbox{20cm}{mA\\mA}\\
\hline
\end{tabular}
    }
    \label{PMT Characteristics3}
\end{table}

% Table IV ----------> Ok
\begin{table}[H]
\captionsetup{font=normalsize, labelfont=bf, labelsep=period}
    \caption{Nominal voltage distribution ratios for JEM-EUSO PMTs where Dynodes from 3 to 10 have voltage distribution ratio 1 \cite{PMThamamatsu}}
    \centering
    \resizebox{8cm}{!} {
\begin{tabular}{l|c|c|c|c|c|c|c|c|c|c|c|}
\hline
Electrodes&K&Dy$_1$&Dy$_2$&Dy$_3$&Dy$_{10}$&Dy$_{11}$&Dy$_{12}$&G.R&P\\
\hline
\hline
Ratio&2.3&1.2&1.2&1&1&1&1&1&0.5\\
\hline
\end{tabular}
    }
    \label{PMT Characteristics4}
%\end{sidewaystable}
\end{table}

\section{PMT Reliability Analysis}

\subsection{PMT Failure Rate}

Failure rate is the frequency with which an engineered system or component fails, expressed, for instance, in failures per hour. It is often denoted by the Greek letter $\lambda$ and is important in reliability engineering. The failure rate of a system usually depends on time, with the rate varying over the life cycle of the system. In practice, the Mean Time To Failure (MTTF) $\frac{1}{\lambda}$, is often reported instead of the failure rate. This is valid and useful if the failure rate may be assumed constant, often it is used for complex units or systems, electronics, and is a general agreement in some reliability standards (Military and Aerospace). Failure rates are important factors in the insurance, finance, commerce and regulatory industries and fundamental to the design of safe systems in a wide variety of applications.

We can classify failures in two types according to their effects on the device: catastrophic and partial failures. 

\subsubsection{Catastrophic Failure}

A catastrophic failure is a sudden and total failure of some system from which recovery is impossible. Catastrophic failures often lead to cascading systems failure. The term is most commonly used for structural failures, but has often been extended to many other disciplines where total and irrecoverable loss occurs. Failures such as cracks in the glass envelope, electrodes no longer connected or short-circuited, resulting in useless or non-existent signals at the output are considered catastrophic failures in PMTs.

\subsubsection{Partial Failures}
Partial failures are fails in a system or part of a system from which recovery of part or all the system is possible. These kind of failures include temporal failures. Failures such as high dark current (noisy tube), gain drifts, decrease of cathode sensitivity, resulting in incorrect or degraded signals at the output are considered partial failures.

Due to long-term stability problem, an aging period has to be applied whenever tubes have not been operated for a long time. PMTs which do not reach a fairly constant gain after that period can be regarded as failing. This clearly indicates that the reliability of PMTs is closely connected to the mode of operation (voltage, mean anode current, etc.) and to the specifications which are expected from these tubes. It is also clear that photomultipliers are more like systems than components as far as reliability is concerned due the numerous characteristics affecting their operation (cathode sensitivity, noise, gain, etc.) \cite{Trends in Photomultipliers}.

The reliability analysis of the PMT is a critical topic in ground-based experiments and space missions where the proper working condition of PMT's are crucial for the success of the experiment. The reliability of the JEM-EUSO PMTs should be high due to the big amount of PMT necessary to implement the whole focal surface of the JEM-EUSO Space telescope. The propagation of failure rate caused by this amount of components makes mandatory the usage of high reliable components. As usual, two types of failure must be determined \cite{Trends in Photomultipliers}.

The main scope of the reliability analysis could be summarized to assess that PMT design satisfies the required quality/reliability objectives and mission specifications and to assess the quality and reliability are maintained at prescribed levels. The analysis consists of three main sections: 

\vspace{-2mm}
\begin{itemize}[]
\item To define the overall purpose, scope , nomenclature, and general reference material for the reliability analysis. 
\item To define the analysis methods used and the calculated failure rates of system components.
\item To state conclusions, including the overall system reliability.
\end{itemize}

\subsection{Rational For Excluded Items}

The following elements are excluded from this assessment:

\begin{itemize}

\item Structural elements, as far as they are controlled by demonstrating positive Margin of Safety when applying Factor of Safety to design strength, in order to warrantee the structural capacity of the system beyond the expected loads. Indeed the Factors of Safety are related to reliability. Design is then qualified and the Hardware Manufacturing tested for acceptance. Fatigue and fracture analysis may be applied to identify resistance versus life time of the mission (in this case 5 years instead of 3 years are required). 

\item Non electronic optical elements, like lenses and filters, shall be demonstrated to meet the required functional  performance by a proper dimensioning at End of Life conditions and the application of life Safety Factors, so that their degradation does not degrade the mission during the required mission lapse of time. Under a structural point of view they shall be treated with the same philosophy of structural items.

\item Passive thermal control, like Multi Layer Insulation (MLI) and thermal washers, shall be demonstrated to meet the required thermal performances by a proper dimensioning at End of Life conditions and the application of Life Safety Factors, so that their degradation does not degrade the mission during the required mission lapse of time.

\item Hardness and connectors are controlled by using qualified materials and by applying the required Safety Factors in the design phase. Manufacturing by qualified providers, integration by verified procedure, verification campaign shall warrantee the required level of reliability. These values will be assessed in a future paper.

\item Software (if any) is controlled by on ground SW Quality Assurance according to the identified level of criticality. Interactions between Hardware and SW might be assessed in HSIA (Hardware-Software Interaction Analysis) in a later stage of the project, to verify the reaction of SW after any HW failure.

\end{itemize}

\subsection{Component Reliability Analysis}

The analysis was conducted for the JEM-EUSO PMT in accordance with the "Parts Stress Analysis" Method given in MIL-HDBK-217-F \cite{milhdbk, 217Plus}.

Following assumptions were made during analysis for each component and the stress analysis:

\begin{itemize}[noitemsep,nolistsep]
\item Uniformity. All elements which compose the PMT must be statistically identical (within the same batch number).
\item Independent trials. The success or failure of any component must not affect the outcome of the next trial.
\item Constant rate of failure. The rate of failure from one trial to the next must remain constant.
\item Space Environment
\item Ambient temperature ($-30^{\circ}$C to $40^{\circ}$C)
\item Junction temperature 125$^{\circ}$C to $150^{\circ}$C
\item Duty Cycle $\sim 20\%$
\item Cycling Rate = 5840 $\sim $  (16 days $\times$ 30 days $\times$ 1 year)
\item Electrical stress $\leq $ 0.5
\end{itemize}

\section{Partial failure evaluation of JEM-EUSO PMTs}

\subsection{Quantum Efficiency Degradation}

Definitely, the lifetime of the PMT is linearly dependent of the working conditions when using it as well as for the charged applied on it. The following equations can explain clearly this situation

\begin{equation}
QE(t)= QE_0\ e^{-t/\tau}
\end{equation}
where $QE_0$ is the quantum efficiency of a new unused PMT;  $\tau$ is the lifetime of the cathode PMT under working conditions. As the PMTs lifetime is depending of the use, $\tau$ is not a good parameter to characterize PMTs by its dependence on the working conditions. 

\begin{equation}
QE(q)= QE_0\ e^{-q/\tau_q}
\end{equation}
where $\tau_q$ is the lifetime of the cathode PMT under charges applied on it. In many cases, this value is roughly 2-3 C/cm$^2$.

However, in case of evaluating the time dependent quantum efficiency, the next relation is assumed because of the linearity of both equations:

\begin{equation}
\frac{t}{\tau} = \frac{q}{\tau_q}
\label{relation}
\end{equation}

Therefore, in order to perform the analysis of the Q(q), the q is necessary to be evaluated, where q, is the charge collected at the cathode, expressed in C/cm$^2$ which is determined as follows:

\begin{equation}
q = \frac{e^- \cdot \phi_\gamma \cdot t \cdot E_{\gamma} \cdot Q_{E_O}}{ Ei \cdot S_{PMT}}
\end{equation}

Since q is the relation between the charge collected per unit area, some considerations have been taken into account in order to express it including the intrinsic/physical parameters of the material. To express the Charge, into couloumbs, for instance it is necessary to consider the effect of the gain as the quality of the material to amplify the photo-current received, and this photocurrent is basically the product of the electron's charge by the flux of photons received as well as by the product of the initial quantum efficiency of the material, which is the amount of electrons produced per photon perceived. The division of the equation is related to the unit area necessary to express the charge collected. For instance, it is necessary to multiply the unit area by the Ionization Energy because it is basically the energy necesary to remove an electron per unit area.

Where the values considered to calculate the charge collected are provided in table \ref{Qe_Characteristics}

\begin{table}[!htb]
\caption{JEM-EUSO PMTs QE parameters} 
\begin{center}
\begin{normalsize}
    \centering
\begin{tabular}{|c|c|}
\hline
\textbf{Parameter} & \textbf{Value} \\
\hline
Initial Quantum Efficiency ($Q_{E_O}$) & 20 \% \\
\hline
Electron's Charge (e$^-$)& 1.6$\times$ 10$^{-19}$C\\
\hline
Photon Flux ($\phi_\gamma$)&1.2$\times$ 10$^{9}$ph/s\\ 
\hline
Time (t)&5 Years\\
\hline
Photons Energy ($E_{\gamma}$ )& 4.14 eV\\
\hline
Charge collection ($\tau_q$) & 2-3 C/cm$^2$\\
\hline
Ionization Energy (Ei) & 1.9eV \\
\hline
Photocathode Active Surface ($S_{PMT}$) & 5.29 cm$^2$\\
\hline
 \end{tabular}
   \label{Qe_Characteristics}
\end{normalsize}
\end{center}
\end{table} 

Anyway, considering equation \ref{relation}, the lifetime can be calculated as follows:

\begin{equation}
\tau = \frac{\tau_q \cdot E_i \cdot S_{PMT}}{e^- \phi_{\gamma} \cdot E_{\gamma} \cdot Q_{E_O} }
\label{lifetime}
\end{equation}

Finally, according to equation \ref{lifetime} , the lifetime is 1.264$\times$ 10$^{11}$s. Therefore, the quantum efficiency over time is as Equation \ref{quantum_time} reflects, considering t as 5 year functioning.

\begin{equation}
Q(t)= Q_{E_O} e^{-t/\tau} = 19.975 \%
\label{quantum_time}
\end{equation}

In other words, a loss of 0.15\% of the quantum efficiency in 5 years per PMT as seen in Figure \ref{qeff} where the first 5 years of degradation are expressed by the red line over the blue line. According to this study, the quantum efficiency degradation is absolutely negligible. It can be also said that it is related to the low level of illumination the PMTs are going to be exposed to during the time of the mission.

% Figure II ------> Ok!
\begin{figure}[H]
\begin{center}
    \centering
\includegraphics[scale=0.125]{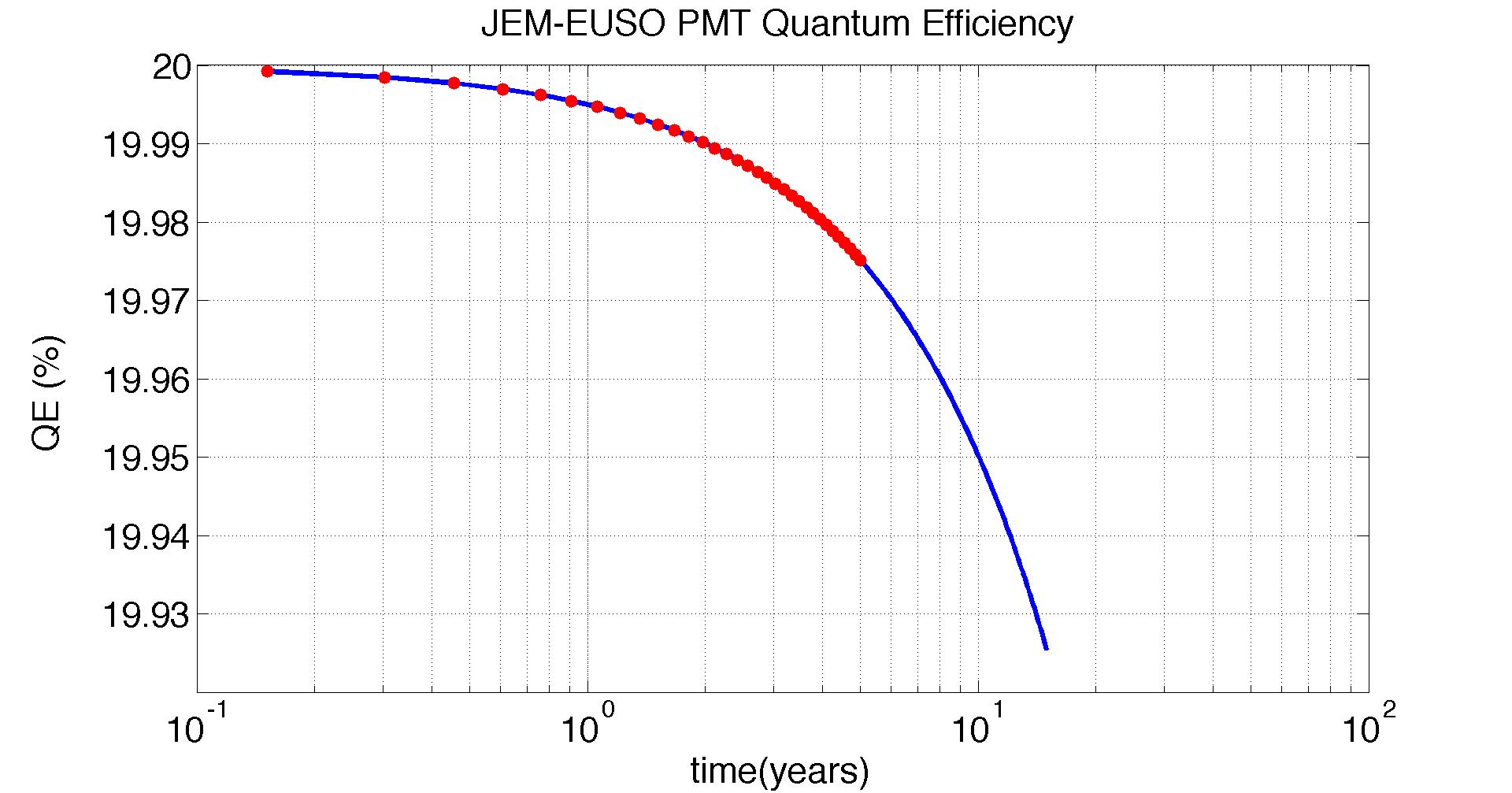}  
\end{center}
\caption{JEM-EUSO PMTs Quantum Efficiency degradation expressed at different time periods. According to this results, the QE degradation is negligible}
\label{qeff}
\end{figure}

Therefore we consider the failure rate of the PMT due to the degradation of the quantum efficiency as the inverse of the lifetime, as $\lambda_{QE}$ = 2.85$\times$ 10$^{-2}$ Failures/10$^6$ h. Hence, the MTTF is 35$\times$10$^6$ h.

Since the Quantum efficiency affects some performance characteristics of the PMT as Sensitivity and Detectivity, their affection and contribution has been evaluated in the following subsections. However, since their failure rate is related to the Quantum efficiency, their value has not been calculated but considered anyway as a contribution to the total failure rate of the QE expressed above. Anyway, its important to higlight that Sensitivity ad Detectivity are expressions of the QE affection, their study it is necessary to understand the meaning of the degradation of the QE. 

\subsection{PMTs Detectivity Degradation}

Noise Equivalent Power (NEP) is the quantity of incident light equal to the intrinsic noise level of a detector. In other words, this is the quantity of incident light when the signal-to-noise ratio (S/N) is 1 \cite{detectivity}.

The inverse value of NEP is the detectivity (detection capability). The detectivity is a measure of the least detectable radiant power or detector signal to noise ratio. A higher D indicates ability to detect lower levels of radiant power \cite{detectivity2}.

Therefore, the NEP is calculated as follows:

\begin{equation}
NEP = \frac{hc}{\eta\lambda}\left(\frac{2I_D}{q}\right)^{1/2} = 1.48\times 10^{-17} W
\label{NEP}
\end{equation}

Where the values considered to calculate the dark current are provided in table \ref{NEP_Characteristics}

\begin{table}[H]
\caption{JEM-EUSO PMTs NEP characteristics} 
\begin{center}
\begin{normalsize}
    \centering
\begin{tabular}{|c|c|}
\hline
\textbf{Parameter} & \textbf{Value} \\
\hline
Planck constant (h) & 6.62$\times$ 10$^{-34}$ J$\cdot$s\\
\hline
Speed of light (c) & 3$\times$ 10$^8$ms$^{-1}$\\  
\hline
Dark Current (I$_D$)& 1.63$\times$ 10$^{-18}$A\\
\hline
Elementary charge of the electron (q) & 1.62$\times$10$^{-19}$ C\\
\hline 
 \end{tabular}
   \label{NEP_Characteristics}
\end{normalsize}
\end{center}
\end{table} 

Hence, $\eta$ is expressed as a time function is calculated according to equation \ref{quantum_time} and dark current I$_D$ is determined as follows:

\begin{equation}
I_D = aAT^2exp\left(\frac{-\phi_0}{kT}\right) = 1.63\times 10^{-18} A
\label{ID}
\end{equation}

The values considered to calculate the dark current and finally obtain the NEP and therfore the detectivity are provided in table \ref{ID_Characteristics}

\begin{table}[H]
\caption{JEM-EUSO PMTs I$_D$ characteristics} 
\begin{center}
\begin{normalsize}
    \centering
\begin{tabular}{|c|c|}
\hline
\textbf{Parameter} & \textbf{Value} \\
\hline
Richardson's constant (a) & 1.2$\times$ 10$^6$ A/(m$^2$ K$^2$)\\
\hline
Absolute temperature (T) in K & 300 K\\  
\hline
Work function ($\phi$)& 1.52eV\\
\hline
Boltzmann's constant (K)&1.38$\times$ 10$^{-23} J/K$\\
\hline
Cathode Area (A) & 5.29 $\times$ 10$^{-4}$cm$^2$ \\
\hline
 \end{tabular}
   \label{ID_Characteristics}
\end{normalsize}
\end{center}
\end{table} 

\begin{figure}[H]
\begin{center}
    \centering
\includegraphics[scale=0.125]{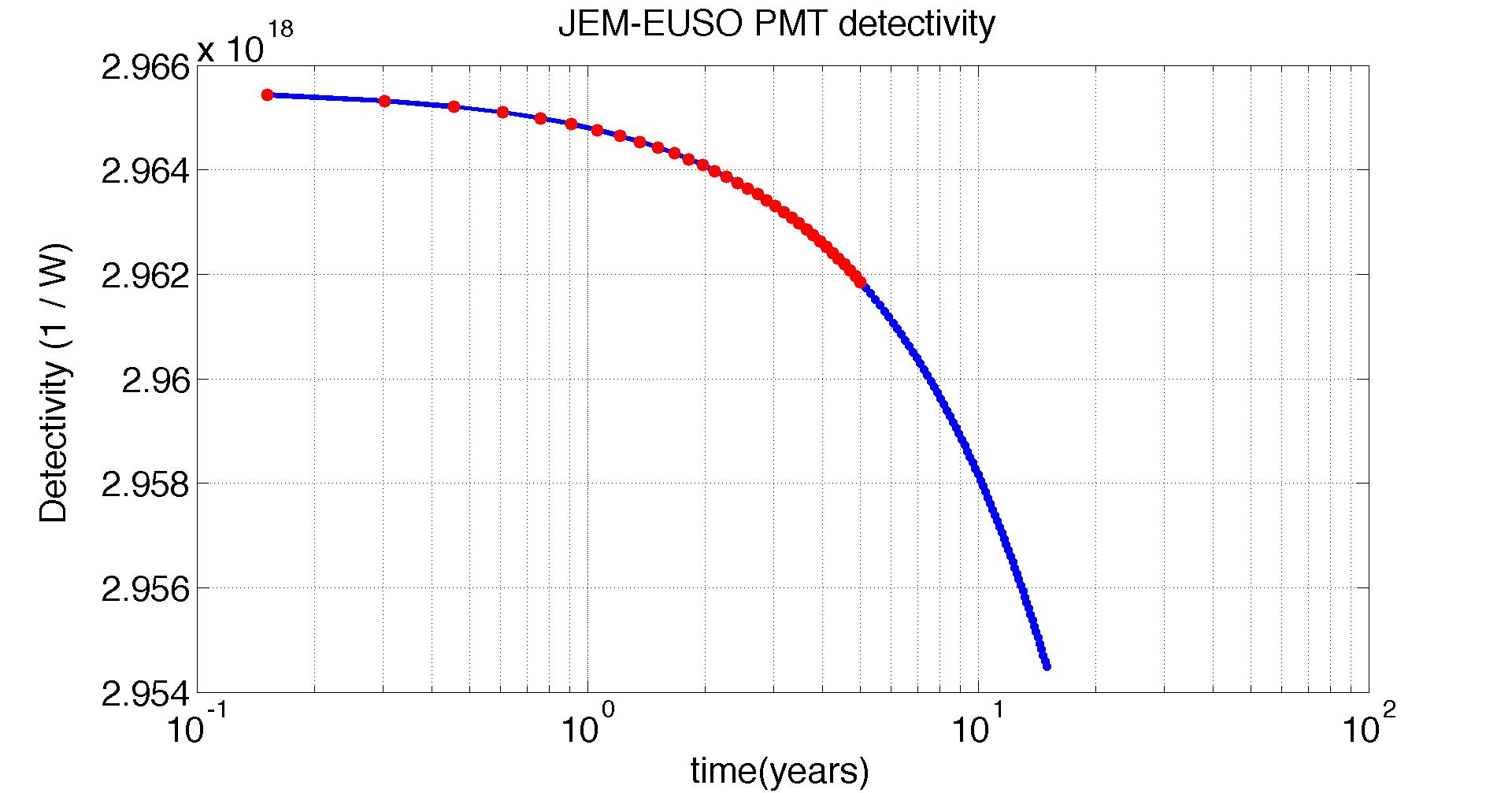}  
\end{center}
\caption{JEM-EUSO PMTs detectivity degradation expressed for different time periods}
\label{detec}
\end{figure}

Previously we mentioned that the reduction of a PMT quantum efficiency is a parameter that directly affects its performance. Hence this can be seen not only as a percentage of the ratio of photons entering the number of subsequently emitted electron. The quantum efficiency degradation can be also express in terms of loss of detectivity of the detector, ie as soon as it has reduced the ability of photons that can be detected throughout the lifespan and therefore the number of photons produced. This effect is shown by Figure \ref{detec}, which clearly shows the linearity between the QE and detectivity. In the end, the reduction of the detectivity is as low as the QE. However its reduction is not considerable enough during the period of the mission since 0.15\% of loss is acceptable considering that 10\% of threshold loss is accepted by the JEM-EUSO science comitee.

% Figure III ------> Ok!

\subsection{PMTs Sensitivity Degradation}

The responsivity expresses how much electrical signal is generated when a given amount of optical flux (power) is incident on a detector, and it can be also consider as the sensitivity of the detector. The electrical quantity can be current or voltage \cite{responsivity}. 

\begin{equation}
R_{esponsivity} = S_{ensitivity} = \frac{\eta q}{h v}
\label{ID}
\end{equation}

Hence, $\eta$ is expressed as a time function is calculated according to equation \ref{quantum_time}, However, the values considered to calculate the sensitivity are provided in table \ref{PMT_sensitivity}

\begin{table}[H]
\caption{JEM-EUSO PMTs sensitivity characteristics} 
\begin{center}
\begin{normalsize}
    \centering
\begin{tabular}{|c|c|}
\hline
\textbf{Parameter} & \textbf{Value} \\
\hline
Quantum Efficiency ($\eta$) & 20\%\\
\hline
Elementary charge of the electron (q) & 1.62$\times$10$^{-19}$ C\\ 
\hline
Planck constant (h) & 6.62$\times$ 10$^{-34}$ J$\cdot$s\\
\hline
Optical frequency (v)&1$\times$ 10$^{15} hz$\\
\hline
 \end{tabular}
   \label{PMT_sensitivity}
\end{normalsize}
\end{center}
\end{table} 

As for detectivity, sensitivity is a parameter related to the quantum efficiency. Its behavior allows to observe the performance of the PMTs. Its understanding also contributes to calibrate the instrument in order to guarantee the success of the mission. However, the sensitivity shown by this study, is within the parameters of the manufacturer, with a value of about 70.49 $\mu$A$/$lm over the first 5 years and a reduction of up to 0.1\% of its value during the same period of time.

% Figure IV ------> Ok!
\begin{figure}[H]
\begin{center}
    \centering
\includegraphics[scale=0.125]{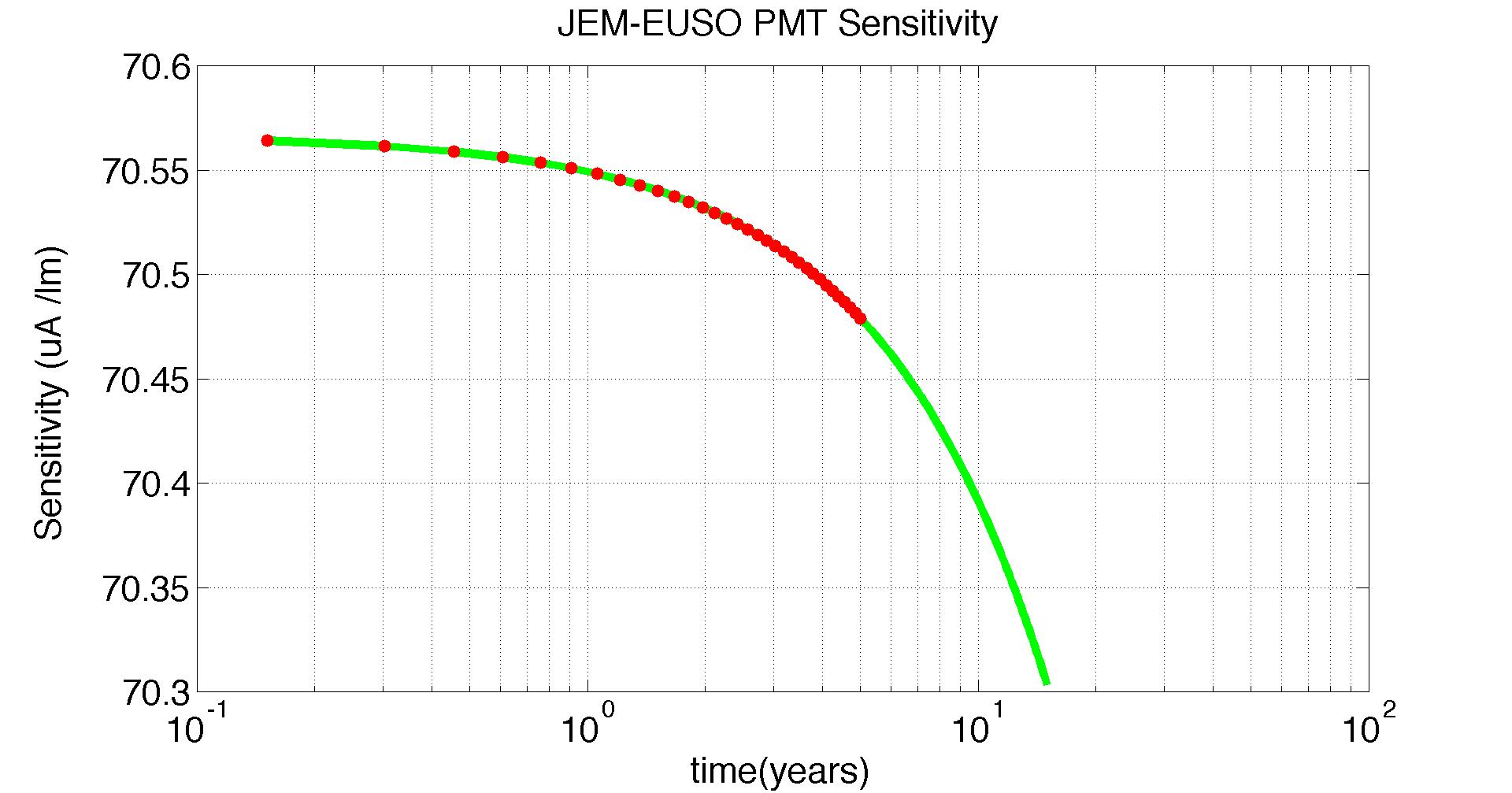}  
\end{center}
\caption{JEM-EUSO PMTs Sensitivity degradation expressed for different time periods}
\label{qeff}
\end{figure}

\subsection{PMTs performance degradation due to radiation}

According to \cite{radiationhp}, the amount of PMTs that will fail during the JEM-EUSO is about 16 including all type of radiation sources. Therefore, the failure rate in terms of radiation is $\lambda_{RAD}$ = 7.3 $\times$ 10$^{-2}$ Failures/ 1$\times$ 10$^{6}$ h, hence, the MTTF of about 1.37 $\times$ 10$^7$. This values will be explained in detail in the nexts subsections when whole sources of failure are considered in order to stablish the total Failure Rate of the JEM-EUSO PMTs.

\section{Catastrophic failure evaluation of JEM-EUSO PMTs using 217 Plus Standard}

Reliability analysis using standard 217 Plus is quite useful since It is basically the latest version of MIL-HDBK-217-F. Also, the 217 Plus has optional data included in order to enhance the predicted failure rate by adding more detailed pertaining  to environmental stresses, operating profile factors and process grades. It also contains default values for the environmental stresses and environmental profile as well as many more and new components missing in MIL-HDBK-217-F. The failure rate ($\lambda_p$) for plastic encapsulated (non-hermetic) integrated circuit applicable for the PMT is given by equation \ref{Failure rate equation for PMT using 217 Plus}, and the  the conditions and values according to the 217 Plus standard are shown in  Table \ref{tab:PMT Condition Values /217Plus} . These values allow us to estimate the base failure rate and failure rate multipliers for any environmental, with operational as well as non operational conditions which the component is exposed to. The calculations results from the failure rate evaluation made to The PMT are shown in Table \ref{tab:PMT Failure Rate Calculations according to 217 Plus}.

\begin{equation}
\begin{split}
\lambda_{P}&=\pi_{G}(\lambda_{OB}\pi_{DCO}\pi_{TO}+\lambda_{EB}\pi_{DCN}\pi_{RHT}+\\
&\lambda_{TCB}\pi_{CR}\pi_{DT})+\lambda_{SJB}\pi_{SJDT}+\lambda_{EOS}
\label{Failure rate equation for PMT using 217 Plus}
\end{split}
\end{equation}

The key factor in failure rate evaluation for PMT according to 217 Plus is: Failure rate multiplier, solder joint delta temperature $(\pi_{SJDT})$, this value reflects that the most important aspect to consider when assembling our electronics system in case of PMT is soldering, a well known critical parameter to be considered, which is now confirmed and reflected by this evaluation.

% Table V ---------> Ok!
%\begin{scriptsize}
\begin{table}[H]
\caption{PMT Condition Values according to 217 Plus}
\begin{center}
    \resizebox{7,9cm}{!} {
\begin{tabular}{|l|c|}
\hline
\textbf{Condition}&\textbf{Value}\\
\hline
Integrated Circuit, Plastic Encapsulated& PMT\\%Ok!
\hline
Year of Manufacture&2010\\%Ok!
\hline
Growth constant ($\beta$)&0.293\\%0k!
\hline
Duty Cycle&20\%\\%Ok!
\hline
Cycling Rate&5840\\%Ok!
\hline
Activation Energy Operating (Eaop)&0.8\\%Ok!
\hline
Activation Energy nonoperating (Eanonop) &0.3\\%Ok!
\hline
Ambient temperature, operating in $^{\circ}$C (TAO)& $26^{\circ}$C\\%Ok!
\hline
Ambient temperature, non operating in $^{\circ}$C (TAE)& $15^{\circ}$C\\%Ok!
\hline
Duty Cycle Operating (DC1op) &0.72\\%Ok!
\hline
Duty Cycle non Operating (DC1nonop)&0.28\\%Ok!
\hline
Default temperature rise ($TR_{default}$) in $^{\circ}$C &$25 ^{\circ}$C\\%Ok!
\hline
Delta Temperature ($DT_{1}$) in $^{\circ}$C &$26.5 ^{\circ}$C\\%Ok%
\hline
Cycling Rate ($CR_{1}$) &482.46\\%Ok!
\hline
\end{tabular}
}
\label{tab:PMT Condition Values /217Plus}
\end{center}
\end{table}
%\end{scriptsize}

% Table VI ---------> Ok!
\begin{table}[H]
\caption{PMT Failure Rate Calculations according to 217 Plus}
\begin{center}
    \resizebox{7.9cm}{!} {
\begin{tabular}{|l|c|c|}
\hline
\textbf{Failure Rate Multiplier (FRM)}&\textbf{Symbol}&\textbf{Result}\\
\hline
Reliability Growth FRM&$\pi_{G}$&$6.867\times10^{-3}$\\%Ok!
\hline
Base failure rate, Operating &$\lambda_{OB}$&$8\times 10^{-6}$\\%Ok!
\hline
FRM for Duty Cycle Operating &$\pi_{DCO}$&0.27\\%Ok!
\hline
FRM for Temperature Operating &$\pi_{TO}$&12.18\\%Ok!
\hline
Base failure rate, Environmental &$\lambda_{EB}$&$1.997\times 10^{-3}$\\%Ok!
\hline
FRM, Duty Cycle, Non operating &$\pi_{DCN}$&1.11\\
\hline
FRM for Temperature environment &$\pi_{RHT}$&0\\%Ok!
\hline
Base failure rate, Temp Cycling &$\lambda_{TCB}$&$8.9\times 10^{-5}$\\%Ok!
\hline
FRM, Cycling Rate & $\pi_{CR}$&12.10\\%Ok!
\hline
FRM,Delta Temperature&$\pi_{DT}$&3.405\\%Ok!
\hline
Base failure rate, Solder Joint&$\lambda_{SJB}$&$4.850\times 10^{-3}$\\%Ok!
\hline
FRM, Solder Joint Delta Temp&$\pi_{SJDT}$&0.635\\%Ok!
\hline
Failure rate Electrical Overstress&$\lambda_{EOS}$&$1.562\times 10^{-3}$\\%Ok!
\hline
\multicolumn{1}{l}{}&\multicolumn{2}{|c|}{Failure Rate = $2.334\times10^{-2}$ F/${10^6}$h }\\  
\cline{2-3}
\end{tabular}
}
\label{tab:PMT Failure Rate Calculations according to 217 Plus}
\end{center}
\end{table}

\section{Estimation of PMTs reliability and Discussions}

Reliability has many connotations. In general, it refers to an item's ability to successfully perform an intended function during a Space mission. The longer the item performs its intended function, the more reliable it is. 

Some systems, such as spacecrafts, cannot be repaired after a major failure, which is obviously the case of JEM-EUSO PMTs. In other cases, even though maintenance tasks can be performed offline, they cannot be performed during a mission. For all of these types of non-repairable systems, the time to system failure is an important reliability characteristic. The expected value is known as mean time to failure (MTTF). \cite{MTTF}.

%% TABLE XIV ----> Ok
\begin{table}[H]
\captionsetup{font=normalsize, labelfont=bf, labelsep=period}
\caption{PMT Failure Rates values}
    \centering
    \resizebox{7.9cm}{!} {
\begin{tabular}{|l|c|c|c|c|c|}\hline
\diaghead{\theadfont Diag ColumnmnHead II}
{Reliability\\Expression}{Study}&\thead{Radiation\\}&
\thead{Quantum Efficiency\\}&\thead{217 Plus \\}\\
\hline
Failure Rate &$7.3\times10^{-2}$ h &$2.85\times10^{-2}$ h &$2.334\times10^{-2}$ h \\
\hline
\end{tabular}
}
\label{pmtmttf}
\end{table}

The failure rates calculated before yield the MTTF Values are listed in Table \ref{pmtmttf}. A MTTF of 8 million hours  (8$\times$10$^6$ hours) according to the addition of every PMT failure source. Certainly they do not mean we can expect an individual device to operate for 913 years before failing. MTTF is a statistical measure, and as such, it cannot predict anything for a single unit. Assuming that during the useful operating life period the PMT have constant failure rates, and part failure rates ($\lambda_{parts}$)  follow an poissonian  distribution. In this case, the MTTF of the product can be calculated as:

\begin{equation}
MTTF= \frac{1}{\sum \lambda_{PMT}}
\end{equation}

and the probability that the product will work for some time t without failure is given by:

\begin{equation}
P(t)=e^{-\frac{t}{MTTF}}
\end{equation}

Thus, in case of our JEM-EUSO PMT with an MTTF of $8\cdot 10^6$ hrs (All failure sources included) and a mission duration of 5 years (43,800 hrs) of continuous operation,

\begin{small}
\begin{eqnarray}
P_{tc}&=&e^{-\frac{43800}{8\times10^6}}=99,45\%
\end{eqnarray}
\end{small}

\begin{figure}[H]
\begin{center}
    \centering
\includegraphics[scale=0.126]{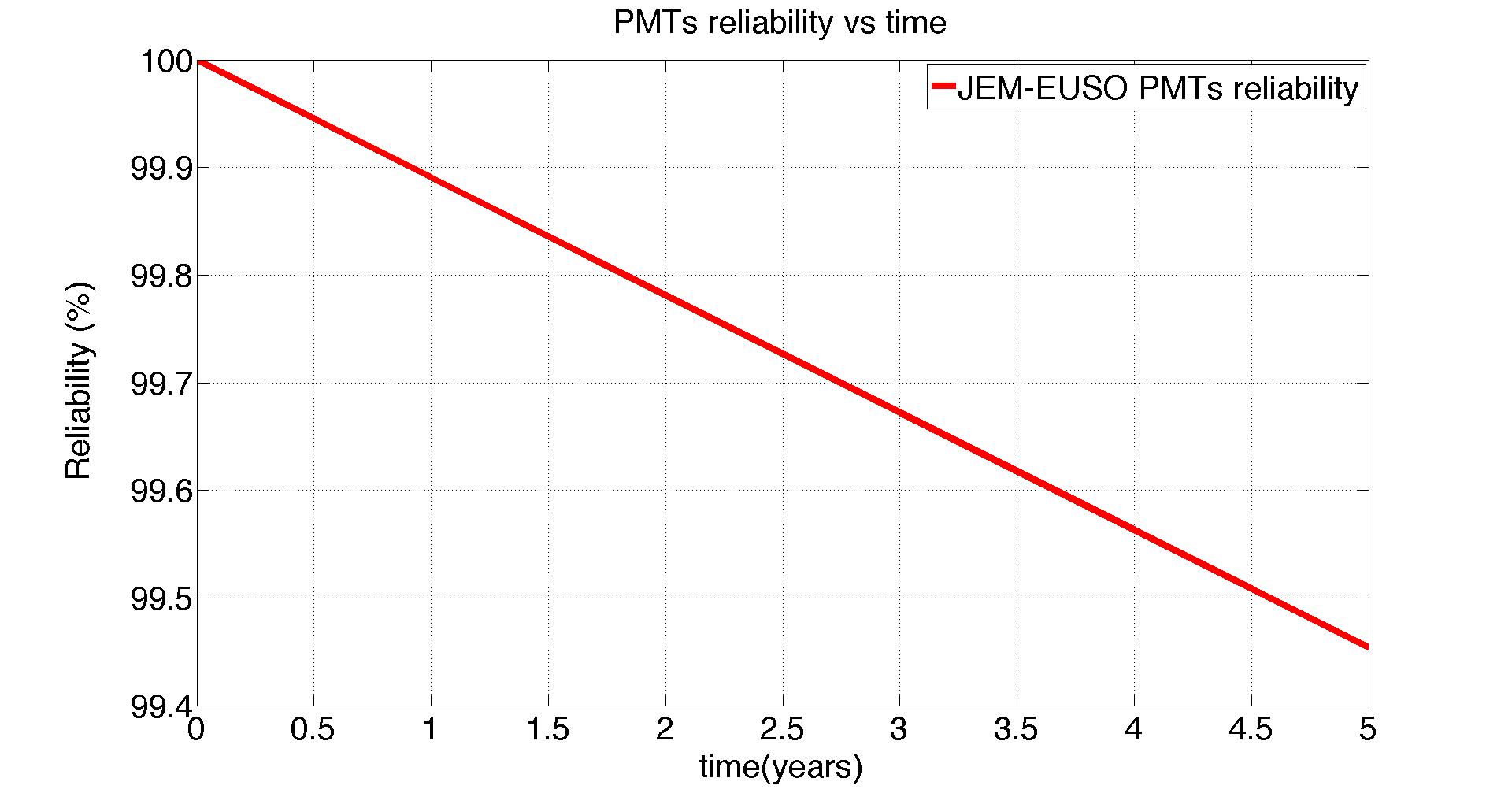} 
\end{center}
\caption{JEM-EUSO PMTs reliability vs time including multiple failure sources}
\label{reli}
\end{figure}

$P_{tc}(t)$ shows that the PMT functioning at the end of 5 years operation period will be operated with 99.45\% of reliability for five years without a failure. Figure \ref{reli} illustrates this situation. In other words, an estimated lost of 0.55\% of the JEM-EUSO telescope focal surface in 5 years of operation period according to 217 Plus, radiation failure sources and Quantum efficiency degradation. We can use MTTF rating more accurately, however, in JEM-EUSO focal surface we have 4,932 PMT operating continuously, so, we can expect 27 PMTs to fail in five operation years according to this study.

% Figure V ------> Ok!

\section{Approximate number of PMTs failures}

In order to have an idea of how the JEM-EUSO PMTs would fail during the time of the mission, it is necessary to determine how this specific failures are going to be distributed. The best way of doing this, is to apply poisson distribution.

The Poisson equation for predicting the probability of a specific number of failures (r) in time (t) is as follows \cite{probability} :

\begin{equation}
P(r)=\frac{(\lambda t)^r e^{-\lambda t}}{r !}
\label{poisson}
\end{equation}

where:

\begin{itemize}[noitemsep,nolistsep]
\item r = number of failures in time (t)
\item $\lambda$ = failure rate per hour
\item t = time expressed in hours
\item P(t) = probability of getting specific r failures in time t
\end{itemize}

\begin{figure}[H]
\centering
\includegraphics[scale=0.12]{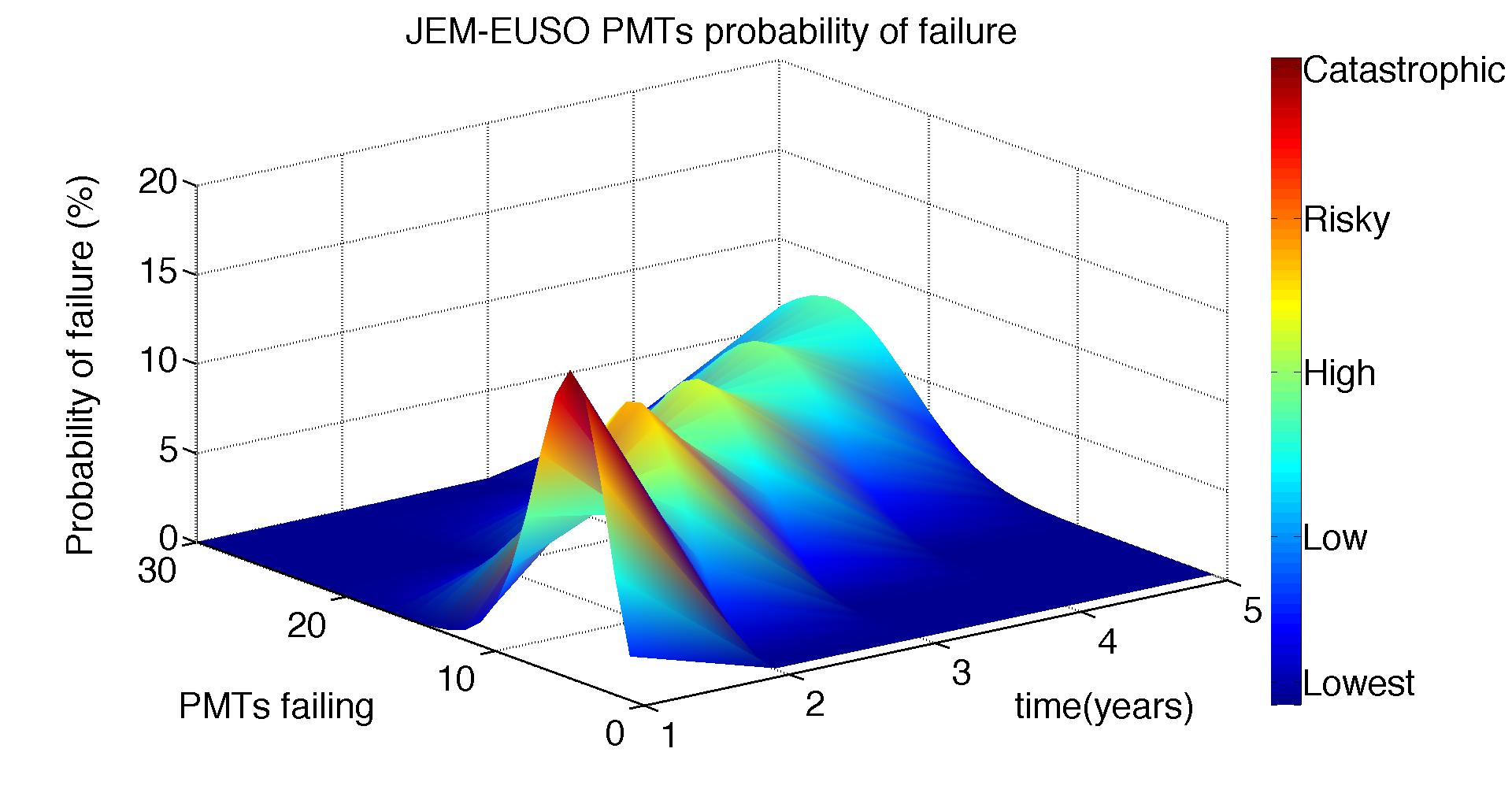}  
\caption{Approximate number of PMT's Failing}
\label{pmtfailing}
\end{figure}

Assuming that the population of the PMT's JEM-EUSO telescope focal surface will use has a failure rate (per year) of 5.4 failures or roughly 27 PMTs are going to fail during the mission due to the quantum efficiency degradation and it is expected to operate during 43800 hours, it is necessary to estimate the especific number of failures during mission time. The values from the estimation were calculated using equation \ref{poisson}, and the results are shown in Figure \ref{pmtfailing}

\section{Conclusions}

Results from this reliability study are as expected. The levels of damage suffered by the PMTs which comprise the focal surface of JEM-EUSO Space Telescope, are acceptable. The results show as well that the greatest contribution to the failure is due to radiation SET, as predicted by \cite{radiationhp}. The guaranteed performance of this equipment is a 99.45\%, an accepted value of reliability thus fulfilling the objectives and technological challenges of JEM-EUSO. It should be noted that the reliability values ​​of the Standard 217Plus, despite being a standard improved, an updated version of MIL-HDBK-217 Plus does not have sections that include the analysis of radiation of space electronic equipment. The recommendation from this study is the inclusion of all real failure effects may face an electronic, mechanical, optical space equipment. 

\section*{Acknowledgments}

The JEM-EUSO Space Mission is funder in Spain under projects AYA2009-06037-E/AYA, AYA-ESP2010-19082 and the coordinated projects AYA-ESP2011-29489-C03 and AYA-ESP2012-39115-C03. We acknowledge CAM ASTROMADRID S2009/ESP-1496 and MICINN MULTIDARK CSD2009-00064 Projects for their finantial support to the Spanish Consortium involved in JEM-EUSO as well. H.Prieto-Alfonso wants to acknowledge the support from International Program Associates (IPAs),which granted one year stay at the RIKEN institute, Wako-shi, Japan.

\end{document}